\documentclass[%
 reprint,%
 amssymb, amsmath,%
 aip,cha,%
floatfix,
]{revtex4-1}
\usepackage{graphics}
\usepackage{amsmath}
\usepackage{amsfonts}
\usepackage{latexsym}
\usepackage{amsbsy}
\usepackage{epstopdf}
\usepackage{bm}%

\usepackage{color}

\usepackage{ulem} %for strikeout "\sout"

\newcommand{\dyadic}[1]{{#1}
\setbox0=\hbox{$\scriptstyle\leftrightarrow$}
   \setbox2=\hbox{$#1$}
   \dimen0=.5\wd0 \advance\dimen0 by-.5\wd2
   \advance\dimen0 by-\wd0
   \kern\dimen0
{^{\hbox{$\scriptstyle\leftrightarrow$}}}}

\begin{document}

\title{Determining the Angle-of-Arrival of an Radio-Frequency Source with a Rydberg Atom-Based Sensor}
\thanks{Publication of the U.S. government, not subject to U.S. copyright.}

\author{Amy K. Robinson}
\author{Nikunjkumar Prajapati}
\affiliation{Depart. of Electr. Engin., University of Colorado, Boulder,~CO~80305,~USA}
\author{Damir Senic}
\affiliation{ANSYS, Inc., Boulder, CO, USA}
\author{Matthew T. Simons}
%\author{Joshua A. Gordon}
\author{Christopher~L.~Holloway}
\email{christopher.holloway@nist.gov}
\affiliation{National Institute of Standards and Technology, Boulder,~CO~80305, USA}

\date{\today}

\begin{abstract}
In this work, we demonstrate the use of a Rydberg atom-based sensor for determining the angle-of-arrival of an incident radio-frequency (RF) wave or signal.  The technique uses electromagnetically induced transparency in Rydberg atomic vapor in conjunction with a heterodyne Rydberg atom-based mixer.  The Rydberg atom mixer measures the phase of the incident RF wave at two different locations inside an atomic vapor cell. The phase difference at these two locations is related to the direction of arrival of the incident RF wave.  To demonstrate this approach, we measure phase differences of an incident 19.18 GHz wave at two locations inside a vapor cell filled with cesium atoms for various incident angles. Comparisons of these measurements to both full-wave simulation and to a plane-wave theoretical model show that these atom-based sub-wavelength phase measurements can be used to determine the angle-of-arrival of an RF field.
%The results demonstrate AoA measurement is possible; however, the accuracy is limited by sources of %field disturbance within and near the vapor cell. Some of these disturbances can be corrected as %discussed in the paper.
\end{abstract}

\maketitle

The ability to measure angle-of-arrival (AoA) is of great importance to radar and advanced communications applications. Here we present a method of determining AoA based on Rydberg-atom sensors. Atom-based sensors have garnered a lot of attention in the past several years because of their many possible advantages over other conventional technologies. Measurement standards  have evolved towards atom-based measurements over the last couple decades; most notably length (m), frequency (Hz), and time (s) standards. Recently  there has been a great interest in extending this to magnetic and electric (E) field sensors. In particular, since the initiation and completion of DARPA’s QuASAR program, NIST and other groups have made great progress in the development of Rydberg atom-based radio-frequency (RF) E-field sensors \cite{gor1, sed1, holl1, holl2, holl3, sed2, tan1, gor2, fan, sim1, sim2, anderson1}.
The Rydberg atom-based sensors now have the capability of measuring amplitude, polarization, and phase of the RF field. As such, various applications are beginning to emerge. These include SI-traceable E-field probes \cite{holl1, holl2}, power-sensors\cite{holl5}, receivers for communication signals (AM/FM modulated and digital phase modulation signals) \cite{song1, meyer1, holl6, cox1, holl4, anderson2}, and even recording musical instruments\cite{holl7}. In this paper, we investigate the capability of a Rydberg atom-based sensor for determining AoA of an incident RF field.

The majority of the work on Rydberg atom-based E-field sensors uses on-resonant electromagnetically induced transparency (EIT) and Autler-Townes (AT) splitting techniques \cite{sed2, holl2, holl3}. The concept uses a vapor of alkali atoms placed in a glass cell (referred to as a ``vapor cell'') as a means of detecting and receiving the RF E-field or signal.  The EIT technique involves using two lasers. One laser (called a ``probe'' laser) is used to monitor the optical response of the medium in the vapor cell and a second laser (called a ``coupling'' laser) is used to establish a coherence in the atomic system. When the RF E-field is applied, it alters the susceptibility of the atomic vapor seen by the probe laser. By detecting the power in the probe laser propagating through the cell, the RF E-field strength can be determined.  This approach has shown to be very successful for determining the magnitude of an RF E-field. However, an alternative approach is required to measure phase, which is necessary to determine AoA.  Recently, we developed a heterodyne technique using a Rydberg atom-based mixer \cite{sim3}. In this approach, a reference RF field is applied to the atoms. This reference RF field is on-resonance with the Rydberg-atom transition, and acts as a local oscillator (LO). The LO field causes the EIT/AT effect in the Rydberg atoms which is used to down-convert a second, co-polarized RF field (referred to as SIG and is the field for which the phase is desired). The SIG field is detuned (by a few kHz) from the LO field. The frequency difference between the LO and the SIG is an intermediate frequency (IF) and the IF is detected by optically probing the Rydberg atoms. This IF is essentially the beat-note between the LO and SIG frequencies. The phase of the IF signal corresponds directly to the relative phase between the LO and SIG signals. In effect, the atoms down-convert the SIG to the IF, and the phase of SIG is obtained by the probe laser propagating through the atomic vapor.

\begin{figure}[!t]
\scalebox{.3}{\includegraphics*{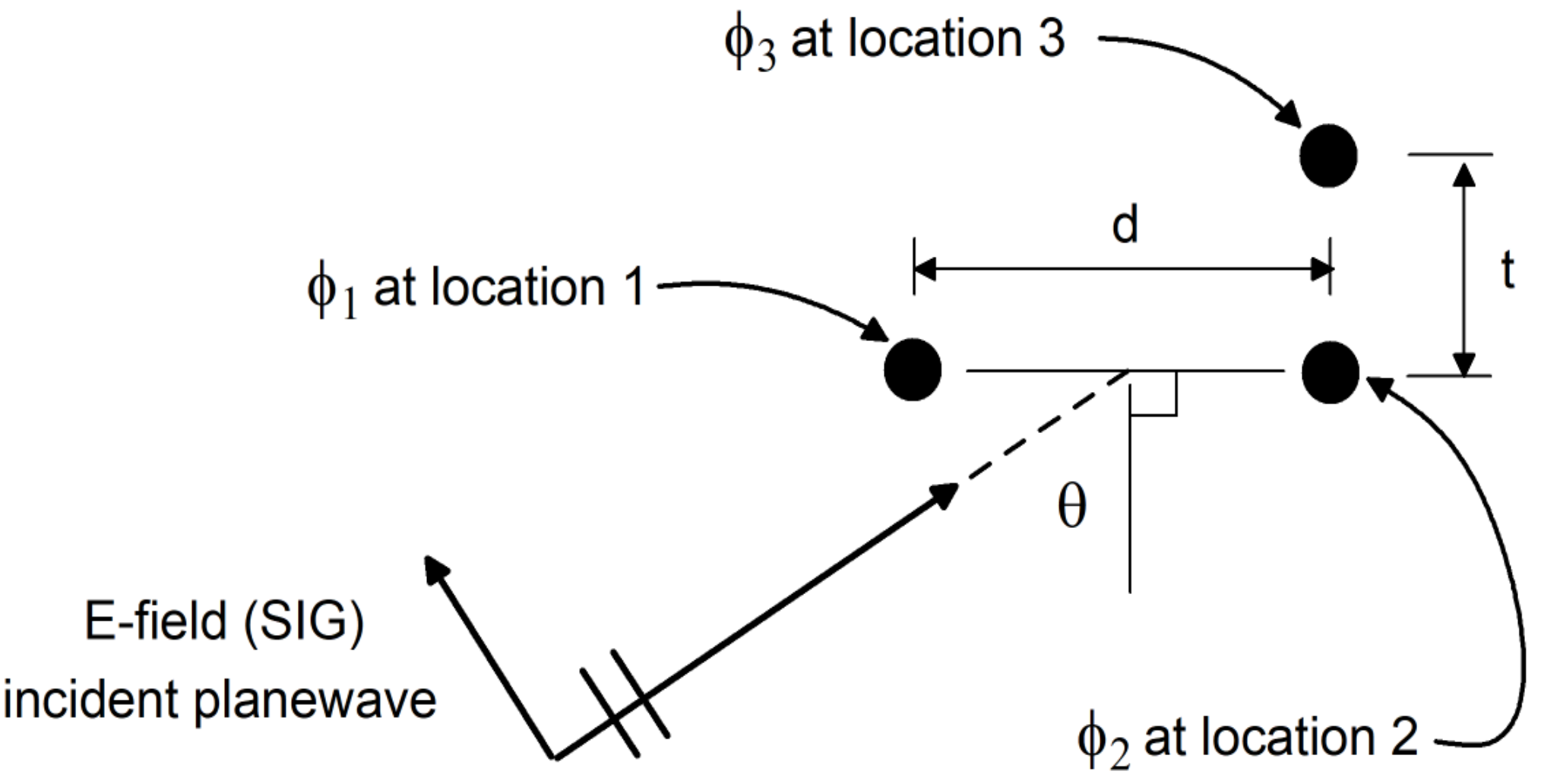}}
\caption{Incident plane wave (SIG) onto three locations separated by $d$ and offset by $t$.}
\label{phasediagram}
\end{figure}

In order to determine the AoA, the phase ($\phi$) of SIG is needed at two different locations, see Fig.~\ref{phasediagram}. Once the phase of SIG is determined at the two different locations, the relationship between AoA (defined as $\theta$ in Fig.~\ref{phasediagram}) and the phase difference at the two locations (location 1 and 2 in the figure) can be calculated. Assuming SIG is a plane wave, the relationship between $\theta$ and $\phi$ is:
\begin{align}
\Delta\phi_{2,1} =\phi_2-\phi_1\approx k\,d\,\,\sin(\theta) :
\theta\approx\sin^{-1}\left(\frac{\Delta\phi_{2,1}}{k\,d}\right)
\label{phase}
\end{align}
where $d$ is the separation between the two locations, $\phi_{1,2}$ are the phases of SIG at the two locations, $k=2\pi/\lambda$, and $\lambda$ is the wavelength of SIG. This expression assumes that the line formed by locations 1 and 2  is perpendicular to the line for which the angle $\theta$ is measured. If the two locations (say locations 1 and 3 in Fig.~\ref{phasediagram}) form a line that is not perpendicular to the line that determine $\theta$, then the phase difference between locations 1 and 3 are given by
\begin{align}
\Delta\phi_{3,1} & =\phi_3-\phi_1 \nonumber\\ &\approx k\sqrt{d^2+t^2}\sin\left[\theta+\tan^{-1}\left(t/d\right)\right]
\end{align}
\begin{align}
\theta\approx\sin^{-1}\left(\frac{\Delta\phi_{3,1}}{k\,\sqrt{d^2+t^2}}\right)\,-\tan^{-1}\left(t/d\right)\,\, ,\label{phase31}
\end{align}
where $t$ is defined in Fig.~\ref{phasediagram}.
Eqs.~(\ref{phase})-(\ref{phase31}) relate AoA to the measured phase of the SIG and LO signals at two locations in the cell, assuming that the AoA is defined in a plane orthogonal to the probe laser propagation. Future work will include the measurement AoA in two dimensions, see discussion below.

\begin{figure}[!t]
\scalebox{.4}{\includegraphics*{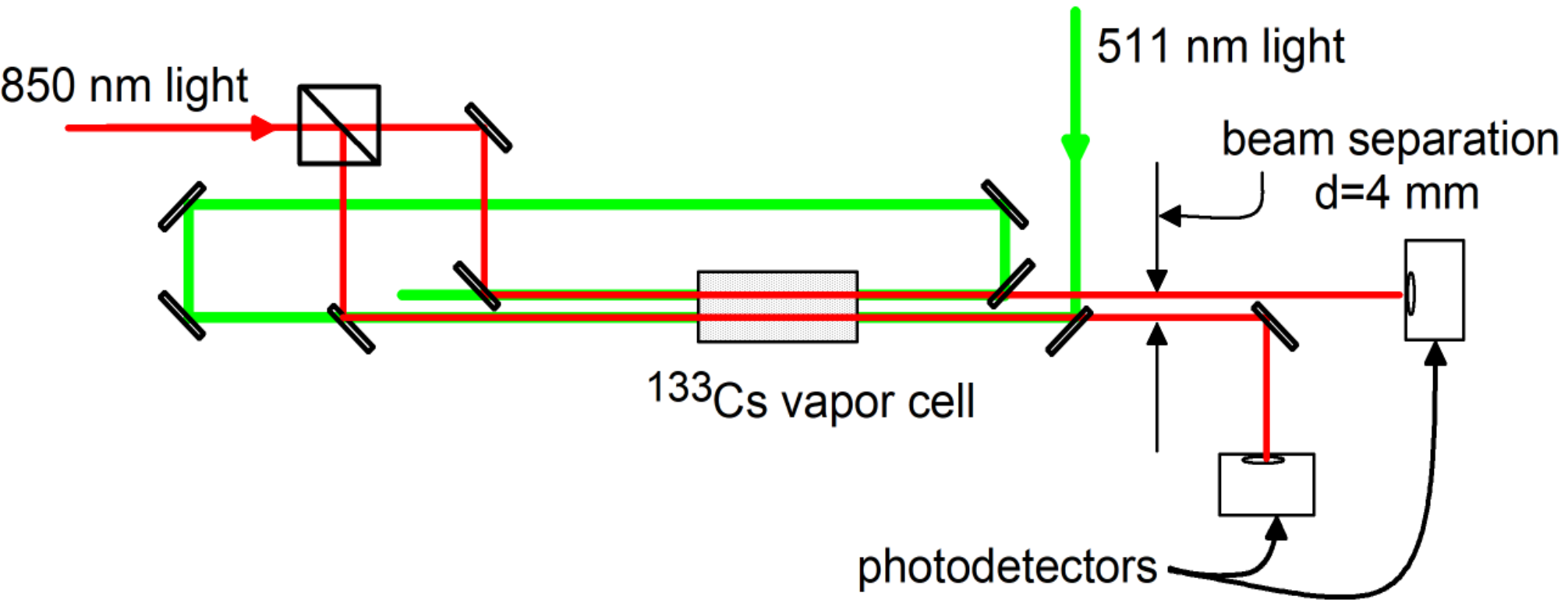}}\\
\begin{centering}
\footnotesize{{(a) Laser field schematic}}\\
%\vspace{2mm}\\
\end{centering}
%\scalebox{.075}{\includegraphics*{LOandSIG.jpg}}\\
\scalebox{.45}{\includegraphics*{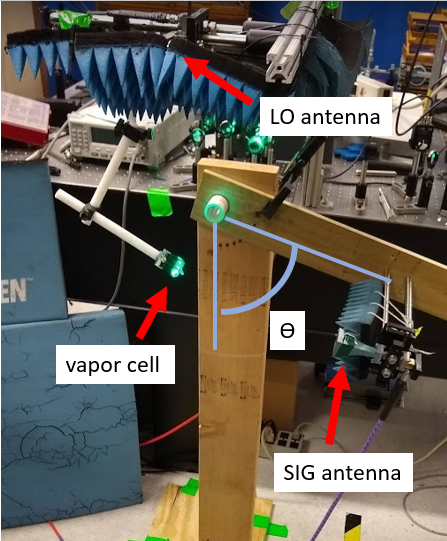}}\\
\begin{centering}
\footnotesize{{(b) antenna arrangement}}\\
\end{centering}
\caption{(a) Schematic of the orientation of the optical fields. The probe beam is split in two by a beam cube, and one coupling field is re-circulated using a dichroic mirror to counter-propagate along each probe beam. (b) The LO antenna is suspended above the cell, such that the LO field is incident nearly perpendicular to the line between the two optical beams. The SIG is held by an adjustable arm to vary the angle of incidence, which is measured using an electronic compass attached to the horn mount.}
\label{setup}
\end{figure}

To measure the phase at any two different locations, we generate EIT in two locations inside a vapor cell filled with $^{133}$Cs, see Fig.~\ref{setup}(a). The probe laser is split with a beam cube and passed through the vapor cell at two locations. The full power of the coupling laser is passed through each of the two locations, see Fig.~\ref{setup}(a). The beam directions are chosen to ensure that at both locations in the cell, the probe and coupling lasers are counter-propagating. To generate EIT at the two locations in the cell, we tune the probe laser to the ${\rm D}_2$ transition for $^{133}$Cs ($6S_{1/2}$-$6P_{3/2}$ or wavelength of $\lambda_p=852.35$~nm) focused to a full-width at half maximum (FWHM) of 390$~\mu$m, with a power of 96~$\mu$W. To produce an EIT signal, we couple to the $^{133}$Cs $6P_{3/2}$-$58S_{1/2}$ states by applying a counter-propagating coupling laser at $\lambda_c=509.26$~nm with a power of 60~mW, focused to a FWHM of 450~$\mu$m.

The LO and SIG are applied to the vapor cell as shown in Fig.~\ref{setup}(b), where the LO is at a fixed position and the SIG is rotated to different incident directions ($\theta$). We use a signal generator (SG) to apply a continuous wave (CW) LO field at 19.18~GHz to couple states $58S_{1/2}$ and $59P_{3/2}$. While we use 19.18~GHz in these experiments, this approach can work at carriers from 100~MHz to 1~THz (because of the broadband nature of the EIT/AT approach \cite{holl1, holl2}).  A second SG is used to generate a CW SIG field at 19.18~GHz+$f_{IF}$ (where the $f_{IF}$=50~kHz)). The output from the two SG are connected to two standard gain horn antennas via RF cables. The LO horn is mounted directly above the vapor cell and is stationary, whereas the SIG horn sits on a rotating arm which sets the incident angle ($\theta$).

\begin{figure}
    \centering
    \scalebox{.15}{\includegraphics*{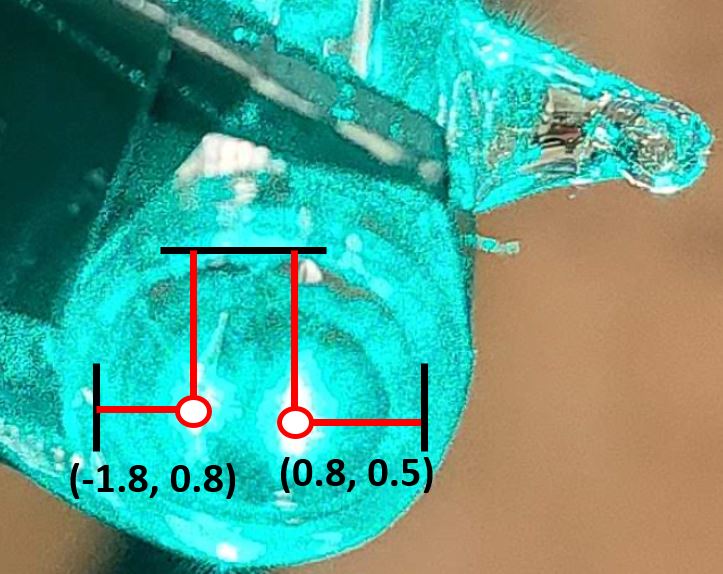}} \hspace{2mm}    \scalebox{.17}{\includegraphics*{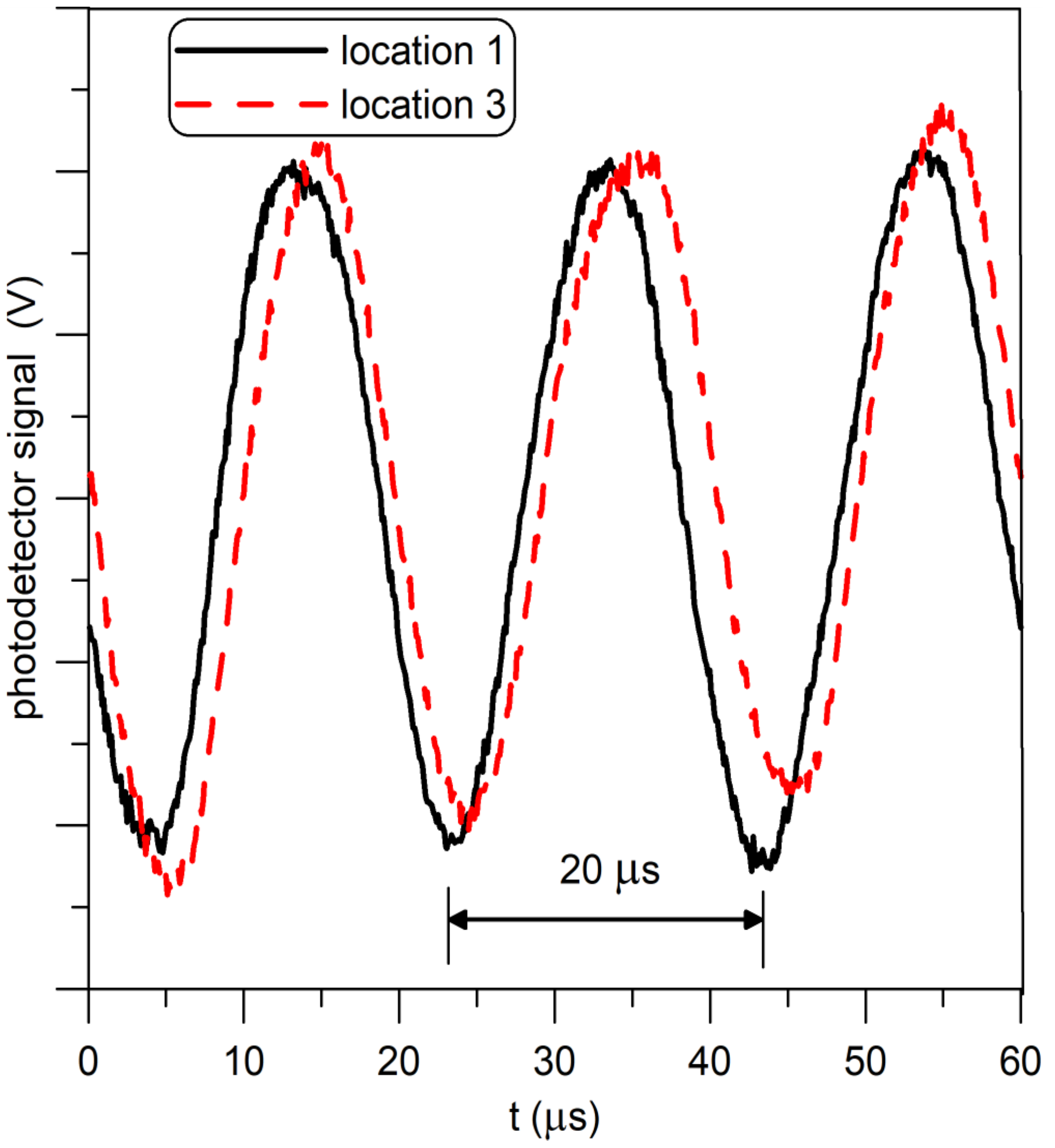}}\\
{\hspace{-1mm}\tiny{(a) \hspace{40mm} (b)}}\\
    \caption{(a)The $x$-$y$ location of lasers inside vapor cell, where the origin is the center of the cell, and (b) Beat-note for two locations inside the vapor cell. }
    \label{cellLasers}
\end{figure}

Two different photodetectors are used to monitor the two probe beams that travel through the vapor cell. The output of the photodetectors are sent to an oscilloscope and a lock-in amplifier.  Fig~\ref{cellLasers}(a) shows the beam position at the two locations inside the vapor cell. These beam positions correspond to locations 1 and 3 as defined in Fig.~\ref{phasediagram}, and the phase relationship is given in eq.~(\ref{phase31}). In our experiments, $d=2.6$~mm and $t=0.3$~mm. The lock-in is referenced to a 50 kHz signal from a mixer that is fed by the two signal generators. The Rydberg atoms automatically down-convert the CW  carrier (i.e., SIG) to the IF (the amplitude of the probe laser transmission) and the phase of SIG is determined.

The Rydberg-atom sensor and the photodetectors act like a mixer and low pass filter in a classic RF heterodyne setup. The LO and SIG create a beat-note and the atoms respond directly to this beat-note, which is detected by the probe laser transmission measured on the photodetectors.  At each location inside the vapor cell, the total electric field ($E_{atoms}$) is the sum of the
LO and SIG fields ($E_{LO}$ and $E_{SIG}$).  The atoms demodulate the high-frequency $\omega_{LO}$ field and the probe
transmission as a function of time at locations $i$ and $j$ (1 and 3 as defined in Fig.~\ref{phasediagram}) is given by~\cite{sim3, gor3}
\begin{equation}
 T_{(i,j)}\propto |E_{atoms}| \approx E_{LO}+E_{SIG}\,\cos\left(\Delta \omega\,t+\phi_{i,j}\right)\,\, ,
\label{Tprobe}
\end{equation}
where $\phi_{i,j}$ corresponds to the phase of SIG at locations i and j, and $\Delta\omega=\omega_{LO}-\omega_{SIG}$.
Once $\phi_{i}$ and $\phi_{j}$ are determined from the probe laser transmissions measured on the two different photodetectors, the phase difference ($\Delta\phi$) between the two locations is given by
\begin{equation}
\Delta\phi=\phi_{j}-\phi_{i} ~~~ \textrm{.}
\label{delta}
\end{equation}
To be more exact,  $\phi_{i,j}$ is actually the phase difference (at each location) between the LO and SIG \cite{sim3}. In these experiments, LO is at a fixed location such that a measurement of $\Delta\phi$ is a measurement of the phase change of SIG between the two locations.

For a given incident angle $\theta$, the beat-notes as measured from the two photodetectors are shown in Fig.~\ref{cellLasers}(b).  From the figure, we see the ``cosine'' behavior as predicted by eq.~(\ref{Tprobe}) with a period of 20~$\mu$s (or the IF frequency of 50~kHz used in the experiments). In this figure we see that the two beat-notes are shifted in phase. This is the phase difference for the given incident angles that is defined in eq.~(\ref{phase31}).

%\begin{figure}[htbp]
%    \centering
%    \scalebox{.20}{\includegraphics*{beatnote_eps.eps}}
%    \caption{Beat-note for two locations inside the vapor cell.}
%    \label{beatnote}
%\end{figure}

\begin{figure}
    \centering
%%    \scalebox{.20}{\includegraphics*{Phase-Angle-Comp_eps.eps}}\hspace{.5mm}
%%        \scalebox{.20}{\includegraphics*{Measure-AoA-CAL_eps.eps}}\\
%%    {\hspace{-1mm}\tiny{(a) \hspace{40mm} (b)}}\\
    \scalebox{.20}{\includegraphics*{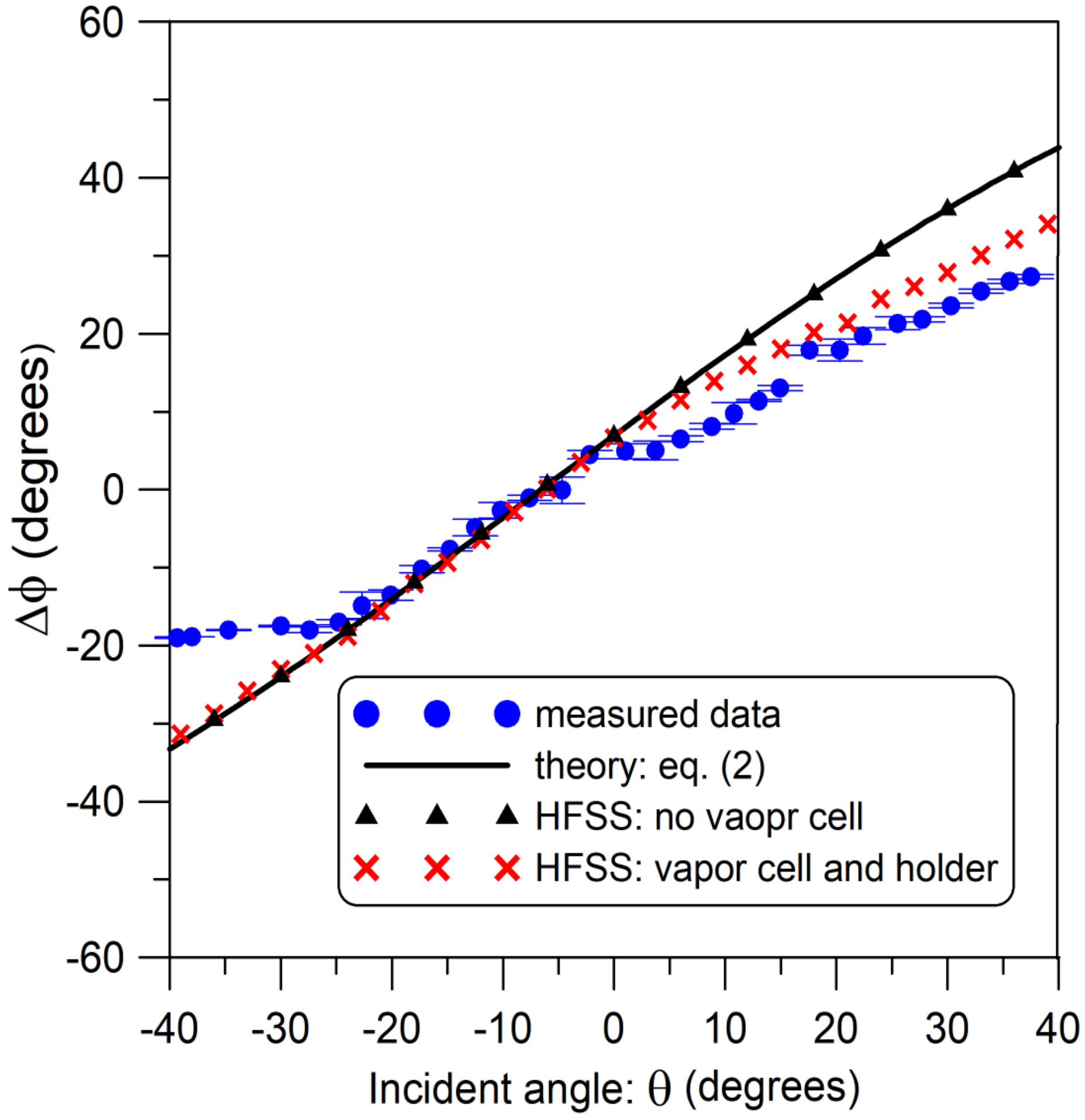}}\hspace{.5mm}
          \scalebox{.20}{\includegraphics*{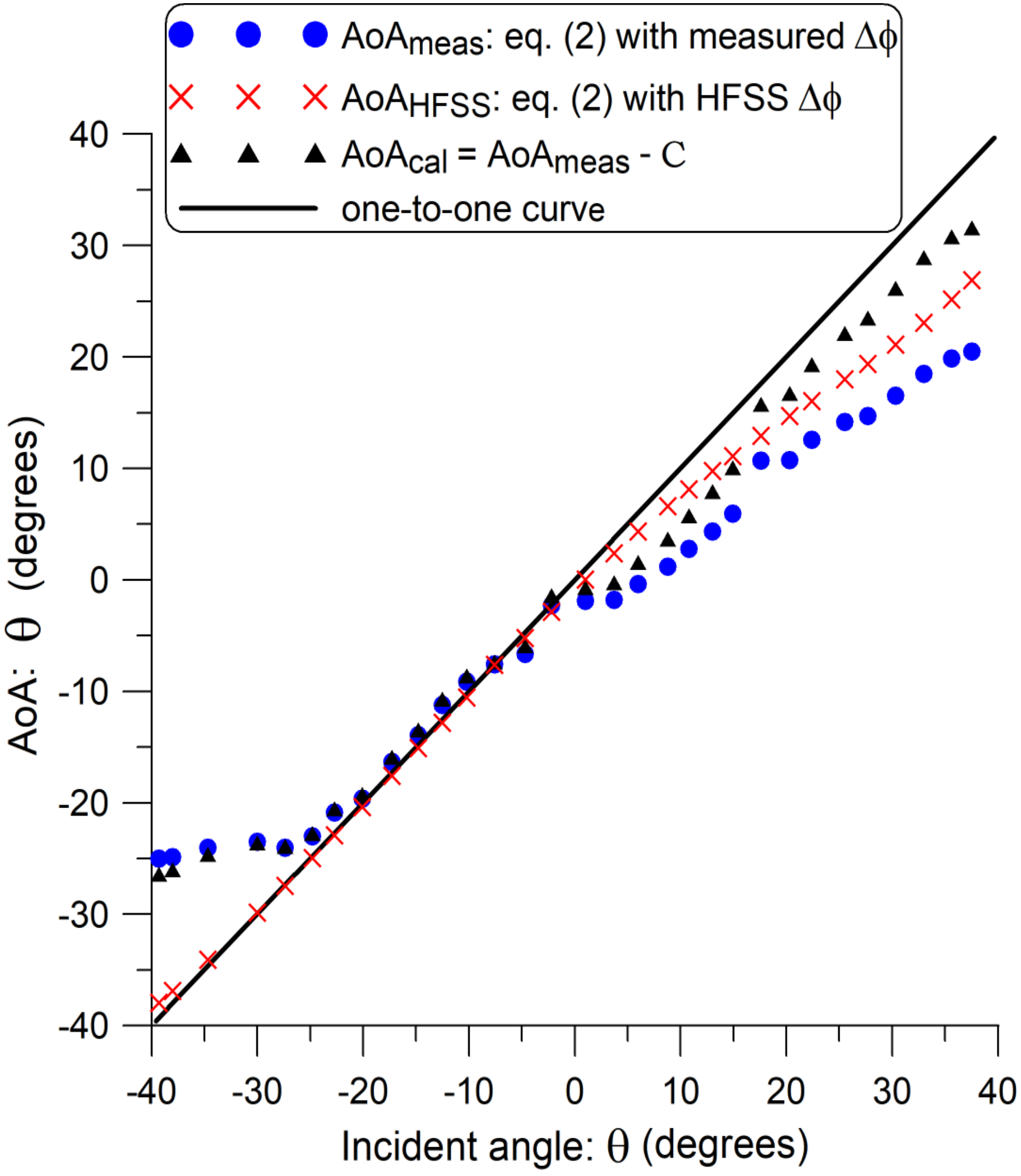}}\\
    {\hspace{-1mm}\tiny{(a) \hspace{40mm} (b)}}\\\vspace{-2mm}
    \caption{(a) Experimental and HFSS data for $\Delta\phi$. The error bars correspond to the standard deviation of 5 data sets, and (b) AoA from the experimental data.}
    \label{compare}
\end{figure}

Using the setup shown in Fig.~\ref{setup}(b), the SIG antenna is scanned from $\theta=\pm40^{o}$. The phase difference ($\Delta\phi$) at each $\theta$ position was determined and the measured $\Delta\phi$ for each incident angle is shown in Fig.~\ref{compare}(a). The error bars correspond to the standard deviation of 5 data sets.  The uncertainties of Rydberg atom based measurements in general are  discussed in Ref.\cite{emcconf} and it is shown in Ref.\cite{ieeeaccess} that the heterodyne Rydberg atom-based mixer approach can measure the phase to within $1^{o}$. Also shown in this figure are the theoretical results given in eq.~(\ref{phase31}). Upon comparing the experimental results to the theoretical results, we see that while the standard deviation for the phase measurement for each incident angle is small (i.e., small error bars), the measurements do not lie exactly on the theoretical results. The reason why the data does not exactly follow the theoretical model is twofold.  First, from Fig.~\ref{setup}(b) we see there are several objects in the apparatus used to rotate the SIG antenna. These objects cause scattering which are not accounted for in the theoretical results. The second reason is due to the vapor cell itself. Because the vapor cell is a dielectric, the RF fields can exhibit multi-reflections inside the cell and RF standing waves (or resonances) in the field strength can develop in the cell\cite{holl2, holl3, sim5, fan5}. Thus, for a given location inside the cell, the RF field can be larger or smaller than the incident field and the phase of the field at a given location will be perturbed as well.  Hence, the standing wave can generates differences in the measured AoA when compared to the expected sinusoidal relationship as given in eqs.~(\ref{phase}) and (\ref{phase31}).
%As the SIG source rotates around the cell, the %standing wave can become increasingly %asymmetrical and larger deviations from the %expected sinusoidal behavior in $\Delta\phi$ %occurs.
Numerical models can be used to investigate this effect. While modeling the entire structure used to support the SIG antenna is difficult, we can use full-wave numerical tools to simulate the vapor cell effects.

\begin{figure}
    \centering
    \scalebox{.28}{\includegraphics*{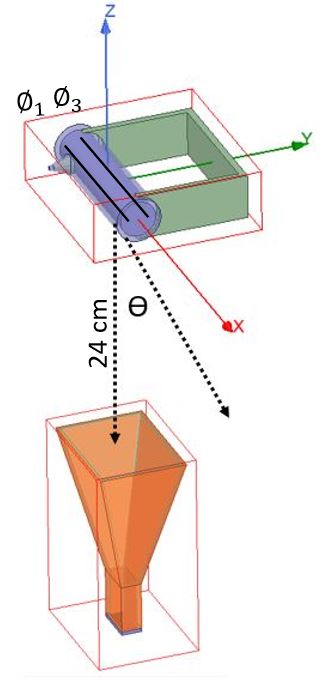}}
    \caption{HFSS model for the cell and horn antenna. The horn is rotated by an angle $\Theta$ around the $x-$axis, such that it points towards the cell. The model also shows the vapor cell holder.}
    \label{horn}
\end{figure}

We use ANSYS HFSS (High Frequency Structure Simulator)\cite{hfss} to simulate only the SIG antenna and the vapor cell (including the plastic vapor-cell holder), see Fig.~\ref{horn}.
%HFSS convergence criteria was defined for both, magnitude and phase, which  %helps with the comparison to measured and theoretical data.
HFSS convergence criteria was based on the energy of a plane wave to 0.01~W, and the mesh around the cell was seeded using curvilinear approximation, and inside the cell using length-restriction to 1~mm, with first order polynomial solving.
With this model, we determine the phase at location 1 and 3 (as defined in Fig.~\ref{phasediagram}) and the $\Delta\phi_{3,1}$ obtained from HFSS are shown in Fig.~\ref{compare}(a). To ensure that the phases are being calculated correctly with the HFSS simulation, we first determine $\Delta\phi_{3,1}$ with no vapor cell present. These results are shown in Fig.~\ref{compare}(a) and match the theoretical calculation closely, as expected. Now that we have confirmed that the HFSS is implemented correctly, the result from the HFSS for the case when the vapor cell is included are shown in Fig.~\ref{compare}(a). We see that the HFSS results (including the vapor cell) correspond well to the measured data for angles \mbox{$>$-25$^{o}$}. As with the experimental results, the HFSS results indicate that the vapor cell does perturb the phase measurement and causes deviation from the theoretical results. We see that the HFSS results do not correspond exactly to the measured data over all the angles, but do show the same trends. The deviations between the measured data and HFSS are twofold. First, the exact permittivity ($\epsilon_r$) of the glass is not known, $\epsilon_r$ ranges from 3 to 6\cite{tropf} (in this numerical model we assume $\epsilon_r=5$). Secondly, upon comparing the photo of the experimental setup in Fig.~\ref{setup}(b) and the HFSS model in Fig.~\ref{horn}, we see that not all the objects used to rotate the SIG antenna are included in the HFSS model. With that said, the measured and HFSS model compare well and follow the same trends, especially for angles $>$-25$^{o}$. There are asymmetries in the apparatus use in the experimenters.
%For angles \mbox{$<$-25$^{o}$} the apparatus used to support the SIG antenna and %vapor cell begins to influence the phase and the apparatus are not accounted for %in the HFSS numerical models.
For angles \mbox{$<$-25$^{o}$} degrees, the apparatus used to support the SIG antenna and vapor cell in the experiment begin to influence the phase of the measured data. The additional scattering caused by this apparatus is not included in the HFSS numerical model.
While the vapor cell does perturb the measurement of $\Delta\phi_{3,1}$, the results in Fig.~\ref{compare}(a) show that the Rydberg-atom based sensor can detect the relative phase difference between two locations inside the vapor cell and work as a AoA detector.
%Rydberg-atom based sensor can detect the relative phase difference between to %locations inside the vapor cell.

With the measured $\Delta\phi$, the AoA can be determined from eq.~(\ref{phase31}).  Fig.~\ref{compare}(b) shows the AoA from the measured $\phi_{3,1}$ for incident angles ranging from $\theta=\pm40^o$.  The solid line in this figure represents the one-to-one correspondence of the incident angle and AoA. The measured AoA should lay on the line. While the measured AoA follows this line, it does not exactly lay on the line. Also, in this figure we show the AoA obtained by using the HFSS results for $\Delta\phi$ in eq.~(\ref{phase31}). Here again, we see that the HFSS results deviate from the solid line. As with the measured $\Delta\phi$, the deviation in the measured AoA (and the HFSS results for AoA) is due to the vapor cell perturbation and due to the supporting apparatus used to experimental equipment (SIG antenna and vapor cell).   This demonstrates that the Rydberg atom-based sensor can be used to determine AoA of an incident RF signal.

%\begin{figure} [h]
%    \centering
%    \scalebox{.22}{\includegraphics*{Measure-AoA-CAL_eps.eps}}
%    \caption{AoA from the experimental data. The error bars correspond to the standard deviation of 5 %data sets.}
%    \label{aoaplot}
%\end{figure}

While the cell does perturb the AoA measurement, two approaches can be pursued to mitigate this effect. One approach is to design a vapor cell that can minimize and even eliminate the vapor cell perturbations. Various groups are investigating different approaches to modify the vapor cell used for these Rydberg atom-based sensors. Two examples include the use of vapor cells with honeycomb sides \cite{schafer} or the use of metamaterials on the sides of the vapor cells \cite{meta}. A second approach is to use the HFSS results to calibrate the vapor cell to reduce the perturbation effects.
This is done by defining a calibration factor as
\begin{equation}
    {\cal C}={\rm AoA}_{HFSS}-{\rm AoA}_{theory}
    \label{cal}
\end{equation}
and subtracting this from the measured AoA
\begin{equation}
    {\rm AoA}_{cal}={\rm AoA}_{meas}-{\cal C}
    \label{aoacal}
\end{equation}
where ${\rm AoA}_{HFSS}$, ${\rm AoA}_{theory}$, and ${\rm AoA}_{meas}$ are the AoA obtained from the HFSS results, theory, and experimental results, respectively. Fig.~\ref{compare}(b) shows ${\rm AoA}_{cal}$. While there is not a perfect correlation to the solid line with the calibration based on the HFSS results, we do see that the calibration did improve the AoA measurement, especially for angles $>$-25$^{o}$. Once again, the deviations from the theory and HFSS simulation for angles $<$-25$^{o}$ is due to the asymmetries in associated with the apparatus to support the experiments. The larger discrepancies could be handled with a more accurate models of the experimental setup.

This paper demonstrates that it is possible to determine the angle of arrival of an RF signal using an atom-based sub-wavelength phase measurement method. While the vapor cell perturbs this measurement, we see that this effect can be mostly accounted for or at least explained, and these Rydberg atom-based sensors have the capability of measuring the AoA of a incident RF signal. Future iterations of this experiment will explore the perimittivity of the glass for more precise modeling of the standing wave in the glass cell. We are also investigating different types of vapor cell designs and beam orientations in order to minimize or eliminate the vapor cell effect on the AoA measurements.

Now that we have demonstrated that it is possible to determine the AoA with a Rydberg atom-based sensor, one can envision (1) developing arrays of these Rydberg atom sensors (2) or sampling the phase at numerous locations inside one vapor cell in order to detect the AoA of more general incidence angles, or for the purpose of simultaneously detecting the AoA of several sources at once.  We are currently developing these two types of sensors and these will be the topic of a future publication.

\end{document}